\documentclass[12pt,preprint]{aastex}
\usepackage{epsfig}

\shorttitle{Numerical simulations of boundary layers}
\shortauthors{J.L.~Fisker \& D.S.~Balsara}
\begin{document}
\bibliographystyle{apj}
\title{Simulating the boundary layer between a white dwarf and its accretion disc}
\author{Jacob Lund Fisker \& Dinshaw S. Balsara}
\affil{Department of Physics, University of Notre Dame, Notre Dame, IN 46556}
\begin{abstract}
We describe the results of numerical simulations of the dynamics of the boundary layer (BL) between the accretion disk and the surface of a non-magnetic white dwarf (WD) for different viscosities which corresponds to different stages for dwarf novae burst cycles.
The simulations cover the inner part of the accretion disk, the BL, and the upper atmosphere of the star.
The high viscosity case, which corresponds to a dwarf nova in outburst, shows a optically thick BL which after one Keplerian rotation period ($t_K$=19s) extends more than 30 degrees to either side of the disk plane. The BL is optically thick and thus occludes part of the star.
The low viscosity case, which corresponds to a dwarf nova in quiescense, also shows a BL, but it is optically thin.
\end{abstract}


\section{Introduction}\label{sec:introduction}
A cataclysmic variable is a close binary system where gas flows from a Roche-lobe overflowing low-mass star via an accretion disk to the surface of a WD \citep{Warner95}. 
This flow releases gravitational energy at a rate of  $L_{acc}=GM_*\dot{M}/R_*$.
In the disk, the gas describes almost circular Keplerian orbits with $v_K\lesssim \sqrt{GM_*/R_*}$.
If viscosity is present in the disk, viscous shear forces between the differentially rotating disk annuli cause friction 
which dissipates energy and turns the orbital energy into heat. 
The heat is radiated away and the resulting energy loss causes the gas to fall into a lower orbit \citep{Shakura73}. 
This is the mechanism behind the accretion process.

In many accreting WD systems, such as dwarf novae and novae-like variables, the magnetic field is weak and the disk extends all the way to the WD surface. 
In order to be accreted, the matter in the accretion disk must lose all of its remaining rotational kinetic energy and decelerate from a nearly Keplerian velocity to the much slower rotational velocity of the surface.
This happens at a rate of $\sim {1 \over 2} \dot{M}v_K^2 \approx {1 \over 2}GM_*\dot{M}/R_*$ which means that about one half of the gravitational energy is released in the BL between the disk and the WD as the rotating matter settles down on the surface while the other half of the gravitational energy drives the disk luminosity \citep{Lynden-Bell74}.
Here the detailed transition of the gas flow determines the relative fractions of how much of the kinetic energy is dissipated as heat and radiated away \citep{Pringle81}, radiated or advected into the surface of the WD, 
and how much is converted into winds, 
and how much goes into spinning up the WD 
and heating it up. 
Additional heat will be released from compressing the atmosphere.
The details of radiated spectrum, therefore, depends on the structure and dynamics of the BL!

The transition region has typically been treated by a BL model, where 
it has been assumed that the averaging of the vertical structure over a disk scale height, $H$ (see \cite{Pringle81}), which is employed in accretion disk studies \citep{Shakura73} can be extrapolated to the surface. 
This reduces the description of the BL to a one dimensional problem \citep{Pringle81, Meyer82, Popham95,Collins98a,Collins98b}.
\cite{Ferland82} pointed out that this radial extrapolation might be invalid.
\cite{Inogamov99} (IS99) presented the first analytic treatment of the latitudinal aspect of the BL.
Averaging over one pressure scale height, $h=P dr/dP$, in the radial direction \cite{Piro04} (PB04) calculated the vertical structure of the BL for a WD and found that the BL could spread towards the poles of the WD.
Early numerical calculations used an axisymetric two-dimensional configuration \citep{Robertson86,Kley91}, 
but since their simulations predated the papers of IS99 and PB04, they did not use enough resolution in the radial direction to capture the dynamics of the BL. 

The objective of the paper is to determine the structural and dynamical dependence of the BL as parametrized by the viscosity. In section \ref{sec:model}, we describe our well-resolved simulations that accurately capture the dynamics of the BL. In section \ref{sec:result} we discuss the multidimensional flows that arise in the BL and present the differences between flows with different accretion rates.

\section{Numerical setup and model}\label{sec:model}
Any numerical simulation must resolve the characteristic scales of the object that is being simulated. 
The scale of the BL is equal to the star's atmospheric scale height in the radial direction and the disk scale height in the vertical direction. 
Furthermore a good model of the BL must include both the inner part of the disk and the outer layer of the atmosphere, so we include five atmospheric scale heights of the stellar atmosphere and three disk scale heights of the disk in our simulation.

To do this we use a spherical mesh with 384 ratioed zones in the radial direction and 128 ratioed zones in the azimuthal range spanning 0 to 30 degrees from the disk plane.
This is sufficient to resolve a BL of a few atmospheric scale heights with more than 100 zones for a very hot star with a temperature of $3.0\times 10^5\textrm{K}$.
The disk has a temperature at the base of $T_0=1.0\times 10^6\textrm{K}$ and it is surrounded by a halo, which is 50 times hotter than $T_0$. The star, disk and halo are tagged by two different species tracers in the simulation.
The toroidal component is assumed to be symmetric around the axis and a reflective boundary condition is imposed at the midplane.
This approximation is justified because the gravitational distortion of the secondary is negligible this close to the primary. 
The equilibrium disk and halo model setup is described in \cite{Balsara04} which builds on models described in \cite{Suchkov94} and \cite{Matsumoto96}. 

We solve the compressible Navier-Stokes equations as given by \cite{Mihalas84} using the \verb+RIEMANN+ code. The inclusion of viscous terms in the momentum equation also requires a self-consistent introduction of viscous dissipation into the energy equation. We follow the prescription of \cite{Mihalas84} in our treatment of viscous energy dissipation.
The spatially and temporally second order agorithms in \verb+RIEMANN+ have been described in \cite{Roe96,Balsara98a,Balsara98b,Balsara99a,Balsara99b} and \cite{Balsara04} and use many ideas from higher order WENO schemes (see \cite{Jiang96} and \cite{Balsara00}) to reduce dissipation.

The matter in the model is subject to the central gravitational field of the underlying WD, which has a radius and mass of $R_*=9\times 10^8\,\textrm{cm}$ and $M_*=0.6M_\odot$ respectively (same as PB04).
The viscosity, which can be molecular, turbulent, magneto-rotational, radiative or a combination thereof, is described by a single parameter, $\alpha$, as in \cite{Shakura73}, where $\nu=\alpha c_s H$ ($c_s$ is the sound speed and $H$ is the vertical scale height of the disk).
In this letter we compare $\alpha=0.1$, which is representative of the disk in the outburst stage \citep{Smak84,Mineshige85} with $\alpha=0.001$ which corresponds to a disk in the quiescent stage.
The model uses an ideal gas ($\gamma=5/3$) of a fully ionized solar composition ($\mu=0.62\,\textrm{g/mole}$) and assumes no radiation transport.

\section{Results and discussion}\label{sec:result}
\begin{figure}
 \epsscale{0.6}
 \plotone{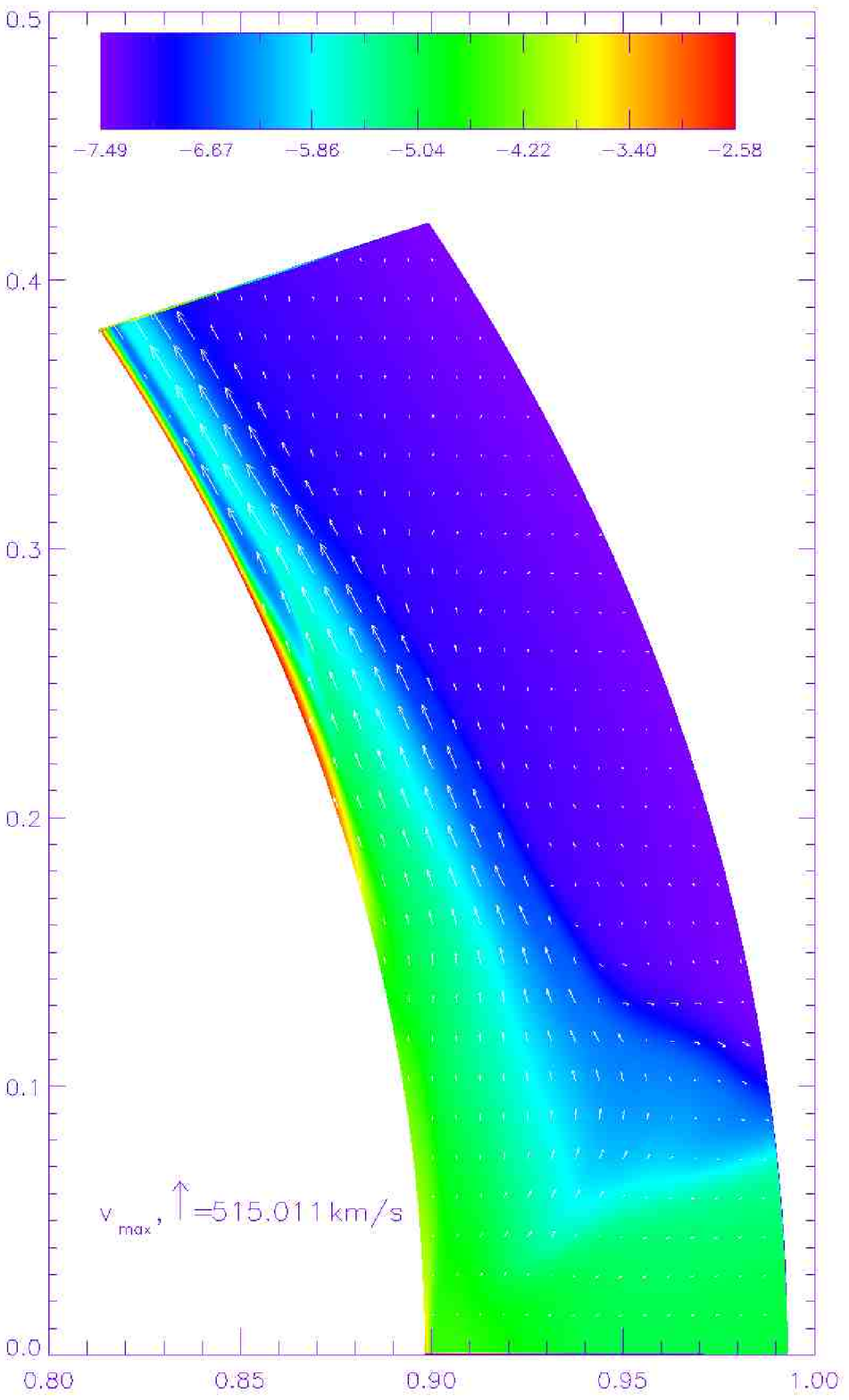}
\vspace{4mm}
 \caption{The plot shows the outermost part of the star (to the left), the BL and the innermost part of the accretion disk (coming in from the right) for the $\alpha=0.1$ case. The baseline of the accretion disk corresponds to the bottom of the plot. The plot only shows the most interesting part of the entire computational domain. The colors show the logarithm of the density in cgs and the axes are in units of $10^9$ cm. Velocity vectors in the poloidal direction are overlaid. The result corresponds to almost one Keplerian rotation period.}\label{fig:densityHIGH}
\end{figure}

\begin{figure}
 \epsscale{0.6}
 \plotone{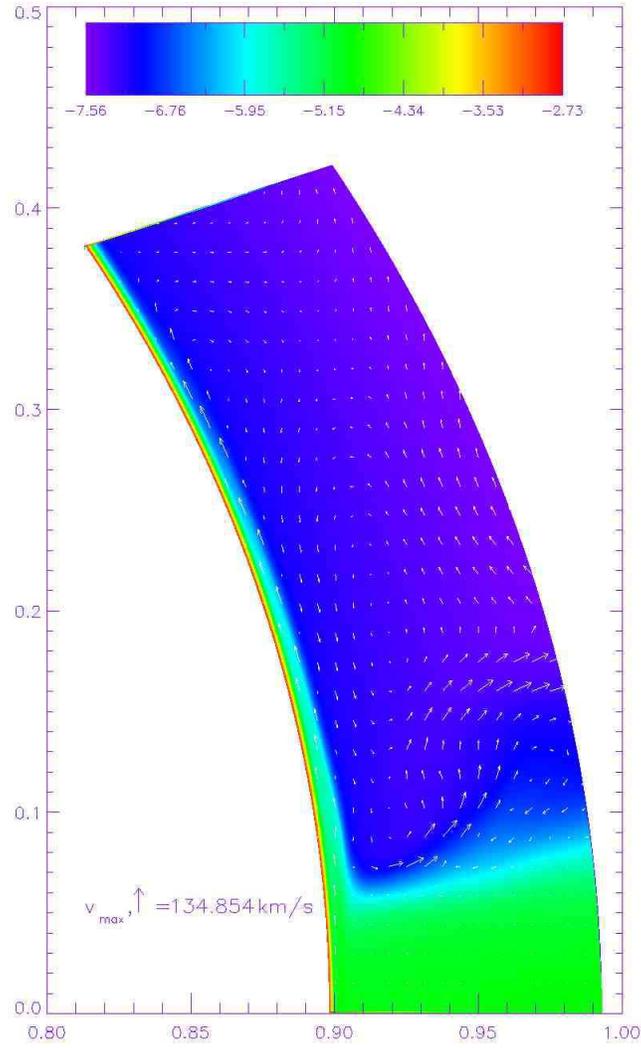}
\vspace{4mm}
 \caption{This plot shows the logarithm of the density in cgs for the $\alpha=0.001$ case at one Keplerian rotation. Notice the absense of waves on the stellar surface and the uniform thickness of the BL.}\label{fig:densityLOW}
\end{figure}

Figs.~\ref{fig:densityHIGH} and \ref{fig:densityLOW} show the logarithm of the density with overlaid poloidal velocity vectors for the $\alpha=0.1$ and $\alpha=0.001$ cases. 
They show that a BL forms at all accretion rates thus corroborating the scenario of IS99 and PB04.
The viscosity has a significant impact on the structure of the BL.

For $\alpha=0.1$, which corresponds to a dwarf novae in the outburst state, the infalling matter depresses the surface and forms a band of disk material that spreads out from the footpoint of the disk.
Here matter keeps piling up as viscous shear forces prevent it from quickly moving to the sides. The shear forces between the rotating band and the underlying star spin up the much denser star matter while slowing down the band.
As the band keeps evolving, friction dissipates energy which heats up the band. 
The pressure in the band increases due to both the increase of internal energy from viscous dissipation and the increase of density from accretion.
At the disk-star interface, the accreting disk matter slows down to very sub-Keplerian velocities while the stellar atmosphere spins up.
At higher latitudes, where the BL is not in contact with the accretion disk, the outer layers of the BL rotate faster than the inner layers. 
With decreasing latitude, the rotation in these outer layers of the BL blend in smoothly with the rotation of the disk.
At lower latitudes, where the BL is in contact with the accretion disk, the pressure and the centrifugal force in the BL prevent the disk matter from accreting radially onto the star.
Thus the pressure and the centrifugal forces in the BL force the disk matter to flow to higher latitudes where the pressure is lower. 
This further enhances the process of forming a multi-dimensional BL. 

In situations with high accretion rates, say $\alpha=0.1$ in Fig.~\ref{fig:densityHIGH}, the ram pressure of the infalling matter can be sufficiently high to depress the surface of the star.
Just as blowing down on the surface of a cup of coffee results in surface waves on the coffee, the depression on the surface of the star causes surface waves to be set up on the star. 
However, since the disk matter piles up too quickly, it soon overflows the surface wave and continues above it towards the pole at supersonic velocities.
This raises the possibility that the poloidal flow is also susceptible to the rapid development of gravitational and/or Kelvin Helmholtz instabilities. Such instabilities have been studied in isolation by \cite{Alexakis04}.
As the overflowing material spills over the surface wave, it creates a substantial amount of friction and heating at the point where it re-establishes strong contact with the star's surface. A hotspot develops at that point of contact, resulting in a reduced density, as can be seen starting at ($r$,$z$)=($0.87$,$0.23$) in Fig.~\ref{fig:densityHIGH}.
This effect is not seen at lower accretion rates, see Fig.~\ref{fig:densityLOW}.

The inclusion of opacity and radiative transport would cause part of the viscously dissipated energy to be radiated away, resulting in a decreased pressure in the BL. Since the pressure contributes to the spreading, it would cause a decreased spreading of the BL. The logical extreme of fully efficient radiative transfer corresponds to a situation where all the dissipated energy is instantly radiated away. This is equivalent to the exclusion of the viscous dissipation terms in the energy equation. We have carried out simulations in this limit also, and we find that the development of the BL is, nevertheless, a robust conclusion. Such simulations will be presented in a subsequent paper.

Eventually the accretion process should settle into thermal equilibrium as energy losses would cause the accreted band to sink and become part of the WD's atmosphere as fast as new matter accretes from above. 
Since we do not include radiative transport we do not trace this thermal equilibrium in the present simulation.
However, for $\alpha=0.1$ (see Fig.~\ref{fig:densityHIGH}) the BL extends more than 30 degrees away from the disk plane and reaches a height of $650\textrm{km}$ ($\sim 0.07R_*$) after close to a Keplerian rotation period. 
The vertical extent of the BL is much larger than the  extent of the inner disc as shown in Fig.~\ref{fig:densityHIGH} Consequently substantial heating of the inner disk by radiation from the BL may change both the density distribution and flow pattern near the inner edge of the disk \citep{Calvet92}.
In this paper we do not focus on the details of spectral modeling. 
Since Thomson scattering will always be present in an ionized plasma, we use it to provide a lower bound on the opacity.
Here we note that the spreading layer would be opaque ($\tau=-\int_\infty^r\kappa\rho dr>2/3$) to Thomson scattering up to at least thirty degrees away from the disk plane on either side. 
For $\alpha=0.001$ (see Fig.~\ref{fig:densityLOW}) the BL expands up to a thickness of about 110 km ($\sim 0.01R_*$) after one Keplerian rotation.
Here, matter simply flows towards the pole in a physically and optically thin layer that is approximately 50 km deep ($\sim 0.006R_*$).

Observations of the UV spectrum of U Geminorum during quiescense \citep{Kiplinger91,Long93,Froning01,Szkody02} show that the WD has a surface temperature of $\sim 30000$--$38000$K. However, additionally there are UV flux levels as well as metal absorption lines corresponding to solar abundances which can only be explained by the additional presence of a small hot ($\sim 57000$K) and slowly decaying continuum source. N V and He II lines, which can only be formed in a hotter ($T>60000$K) and optically thin region, are also observed during quiescense. 
We suggest that these observations may be explained by the transition from the optically and physically thick BL of the outburst stage (Fig.~\ref{fig:densityHIGH}) to the optically and physically thin BL corresponding to the quiescent stage (Fig.~\ref{fig:densityLOW}).

In conclusion, we have verified that BLs are multi-dimensional as suggested by IS99 and PB04 and analyzed the multi-dimensional hydrodynamics of the BL for high and low accretion rates. 
These BLs cover a significant fraction of the surface of the star. 
For high rates of accretion the BL can be optically thick enough to occlude parts of the stellar surface. 

\acknowledgements
We thank L. Bildsten, C. Fryer, A. Piro, E. Sion, D. Townsley, and J. Truran for interesting discussions. JLF was supported by NSF-PFC grant PHY02-16783 through the Joint Institute of Nuclear Astrophysics\footnote{see \email{http://www.JINAweb.org}}. DSB was supported by NSF AST-0132246 and DMS-0204640.
This work used the HPCC cluster at the University of Notre Dame and was partially supported by the National Center for Supercomputing Applications under AST050031 and utilized the NCSA SGI Altix.


\clearpage

\clearpage

\end{document}